\documentclass[10pt,english,10pp]{article}
\usepackage{mathptmx}

\usepackage[T1]{fontenc}
\usepackage[latin9]{inputenc}
\usepackage{amsmath}
\usepackage{esint}
\PassOptionsToPackage{normalem}{ulem}
\usepackage{ulem}

\makeatletter
\newcommand{\lyxaddress}[1]{
\par {\raggedright #1
\vspace{1.4em}
\noindent\par}
}

\renewcommand\[{\begin{equation}}
\renewcommand\]{\end{equation}}

\makeatother

\usepackage{babel}
\begin{document}

\title{A perturbation basis for calculating NMR Diffusometry}

\author{Matias Nordin}

\maketitle

\lyxaddress{Applied Surface Chemistry, Department of Chemical and Biological
Engineering, Chalmers University of Technology, 41296 Gothenburg,
Sweden}

\lyxaddress{\inputencoding{latin1}matias@chalmers.se}
\selectlanguage{english}%
\begin{abstract}
An approximative method for solving the Bloch-Torrey equation in general
porous media is presented. The method expand the boundaries defining
the porous media using electrostatic charges. As a result the eigenvalue
problem of the Laplace operator in a confined geometry can approximately
solved. Importantly the approximative solution is orthogonal in the
low-frequent region of Fourier space. This gives a natural approach
for studying spin magnetization in presence of magnetic fields. The
error in the approximation scales with $N^{-2}$ times the magnitude
of each eigenvalue, where $N$ is the size of the expansion matrix.
From a computational point of view, the calculations scale quadratically
with the number of basis functions using fast multipole methods.
\end{abstract}
Nuclear Magnetic Resonance (NMR) provide an excellent tool for directly
studying transport properties in porous media as well as indirectly
study the porous media itself~\cite{Stejskal1965a,callaghan1991book,price2009}.
Geometrical properties such as surface-to-volume ratio~\cite{Mitra1992,Mitra1993,Hurlimann1994},
characteristic length scales~\cite{callaghan1991,topgaard2003,Malmborg2006}
and pore size distributions~\cite{Price2003,Latour1995,Mair2000,Mair2002}
has successfully been derived. The analysis of the experimental signal
is however in many cases difficult as ~\cite{Grebenkov,Grebenkov2007}.
The main reason is that a theoretical analysis of the NMR experiment
in porous media is difficult~\cite{Grebenkov2007}. A diffusing spin
in a porous media is described by the Bloch-Torrey equation~\cite{torrey1956}.
Analytic solutions exist only for a few simple geometries and for
general media clues may be found from numerical simulations. Another
reason for a difficulty in analysing the experiments lie in the fact
that the experimental conditions for the theoretical models available
cannot always be met, e.g. that the time for the gradient is short
enough for the so-called short gradient pulse limit to hold~\cite{Nordin2011,linse1995,Price2003}.
Therefore, an analysis of the experimental signal is often done using
clues from simple geometries crude models. It is possible to view
this experiment as probing a porous media with low-frequent Fourier
modes, and study the response. In the SGP-limit this aspect is in
fact exact, as the experimental signal is the Fourier transform of
the diffusion propagator in a porous media. If no boundaries are present,
the Fourier modes stay orthogonal, and the experimental signal is
just the exponent of the fourier wave number times the experimental
time. If however the media is not free, the Fourier modes get mixed
as they pass through the media, and the experimental response show
this mixing. Therefore it would be advantegeous to study this mixing,
as it reveal information about the geometry of the porous media.

In this paper an approximate method is presented that connects a general
porous media with the associated eigenspectrum of the Laplace operator
and the magnetic resonance experiment. Importantly this gives the
possibility of directly studying the geometrical impact on NMR diffusometry
experiments in more complex porous media.  

The Bloch-Torrey equation describes a diffusing spin in a porous media
subject to an external magnetic field. For simplicity we assume that
the self-diffusion is isotropic and the following equation is obtained
{[}ref price, Barzykin{]}

\begin{equation}
\begin{cases}
\dot{m}(r,t)=(D_{0}\Delta+i\gamma f(t)G(r))m & r\in\Omega\\
(a\frac{\partial}{\partial n}+b)m(r,t)=0 & r\in\Gamma
\end{cases}\label{eq:original_problem}
\end{equation}
where $D_{0}$ denote the self diffusion coefficient, $\gamma$ the
gyromagnetic ratio and $f(t)$ a time profile of the gradient and
$G(r)$ a magnetic field gradient which can include also internal
gradients. The sought complex valued function $m$ describes the magnetization.
In the case of a freely diffusing spin, the diffusion can easily be
solved by diagonalizing the heat kernel and the resulting effect of
the spin for a given diffusion time $t$ will be dominated by the
eigenfunctions corresponding to the eigenvalues smaller than $e^{-t\lambda_{n}}$
for some largest $n$. In other words, the diffusive motion a spin
undergoes during time $t$ will be dominated by the low-frequent eigenfunctions
up to a certain truncation. The shorter time, the more functions are
needed. It is tempting to think that a similar analysis can be made
also in the case of a spin diffusing in a porous media. This equation
can be solved by formally integrating the magnetization in time (see
e.g. \cite{Barzykin1999}) and expressing the solution in the eigenbasis
of the Laplace operator.
\[
m(r,t)=e^{-t(D_{0}\Delta+i\gamma G(r))}m(r,t=0).
\]
It can also be solved by exploring the solution in Fourier space~\cite{Kenkre1997}
\[
\frac{\partial\hat{m}(q,t)}{\partial t}=\gamma f(t)FT[iG(r)]-D_{0}q^{2}\hat{m}(q,t).
\]
This approach is appealing, as in absence of boundaries the Fourier
modes satisfy the Laplace operator. The problem here is however that
when boundaries are present, the Fourier transformed magnetization
is difficult to solve. One may however note that if such a transformation
is found, the averaged signal is found by the limit~\cite{Kenkre1997}
\[
m(t)=\lim_{q\rightarrow0}\hat{m}(q,t).
\]

The problem stated in Eq. \ref{eq:original_problem} can be solved
by calculating the eigenfunctions and the eigenvalues to the Laplace
operator in the confined domain $\Omega$. This can be stated as
\begin{equation}
\begin{cases}
\Delta u_{n}(r)=\lambda_{n}u_{n}(r) & r\in\Omega\\
(\frac{\partial}{\partial\hat{n}}+a)u_{n}=0 & r\in\Gamma
\end{cases}\label{eq:eigenequation}
\end{equation}
where the porous material is defined by the boundary conditions at
$\Gamma$. 

Let us define the following integral operators to represent the boundary
conditions. For Dirichlet conditions at $\Gamma$
\[
af\mid{}_{\Gamma}=\intop_{\Gamma}\delta(r-r_{0})f(r_{0})dr_{0}=\begin{cases}
af(r) & \mbox{if }r\in\Gamma\\
0 & \mbox{if }r\in\Omega\backslash\Gamma
\end{cases}
\]
where $\delta$ denote the Dirac-delta function. The corresponding
operator for Neuman conditions is defined as
\[
\frac{\partial}{\partial\hat{n}}f\mid_{\Gamma}=\intop_{\Gamma}(\hat{n}(r_{0})\cdot\nabla)\delta(r-r_{0})f(r_{0})dr_{0}=\begin{cases}
\hat{n}\cdot\nabla f(r) & \mbox{if }r\in\Gamma\\
0 & \mbox{if }r\in\Omega\backslash\Gamma
\end{cases}
\]
where $(\hat{n}(r_{0})\cdot\nabla)\delta(r-r_{0})$ denote the distributional
derivative of the Dirac-delta function directed along the normal of
the boundary $\Gamma$ at $r_{0}$ . Using these, we can define the
following inner product for two functions $f$and $g$ 
\begin{equation}
\langle f|\frac{\partial}{\partial\hat{n}}+a|g\rangle_{\Omega}=\intop_{\Omega}f(r)\intop_{\Gamma}(\nabla\cdot\hat{n}(r')\delta(r-r')+a\delta(r-r'))g(r')dr'dr=\label{eq:inner_products_boundaries}
\end{equation}
\[
=\intop_{\Gamma}f(r)(\hat{n}(r)\cdot\nabla+a)g(r)dr
\]
which we may note is zero if the function $g(r)$ satisfies the boundary
conditions. Note that if a function $u(r)$ satisfies Eq. \ref{eq:inner_products_boundaries}
it is also satisfying the boundary conditions in Eq.  \ref{eq:eigenequation}
and can be written as a linear combination of the sought eigenfunctions
$\{u_{n}\}_{n=1}^{\infty}$ in Eq. \ref{eq:eigenequation}. 

In absence of the internal boundary conditions $\Gamma$ a general
function $f(r\in\Omega)$ can be written as a Linear combination of
Fourier modes $f(r\in\Gamma)=\sum_{n=1}^{\infty}\alpha_{n}|q\rangle$.
These Fourier modes satisfy Eq. \ref{eq:eigenequation} \emph{without}
the boundary conditions at $\Gamma$. This can be written as 
\[
\Delta|q\rangle=\lambda_{q}|q\rangle
\]
where the eigenvalues $\lambda_{q}$ are analytic and given by the
exterior boundary conditions (Dirichlet, Neumann or periodic) at $\Omega$.
In example for the case of Dirichlet exterior boundary conditions
with $\Omega$ being a box of side length $L$ the functions $|q\rangle$
equal
\[
|q\rangle=A_{q}\sin(\frac{n_{x}\pi}{L}x)\sin(\frac{n_{y}\pi}{L}y)\sin(\frac{n_{z}\pi}{L}z)
\]
where $A_{q}$ is the normalization constant. The corresponding eigenvalues
equal
\[
\lambda_{q}=\frac{\pi^{2}}{L^{2}}(n_{x}^{2}+n_{y}^{2}+n_{z}^{2}).
\]
We expect that it is possible to expand each eigenfunction of Eq.
\ref{eq:eigenequation} using the set $\{[q\rangle\}_{q=1}^{\infty}$
i.e. Fourier transforming the eigenfunctions. The problem with this
approach is that the eigenfunctions satisfying Eq. \ref{eq:eigenequation}
are not local in Fourier space, meaning that a perturbation expansion
of the eigenfunctions into the set $\{[q\rangle\}_{q=1}^{\infty}$
give poor convergence. The reason is that there is a reciproc relationship
between the boundaries at $\Gamma$ in Eq. \ref{eq:eigenequation}
and the Fourier space on $\Omega$. It would however be of great advantage
if one could find the low frequent behaviour of the solution to Eq.
\ref{eq:eigenequation} in Fourier space as this set has an intimite
relationship with the calculation of the magnetization in NMR diffusometry
(explained below). 

It is a well-known fact (see e.g.~\cite{Amini1999,Bonnet1998,Chen2007})
that given an eigenvalue $\lambda_{n}$ to Eq. \ref{eq:eigenequation}
there exists a corresponding surface distribution $f_{n}(r\in\Gamma)$
such that the following inhomogeneous Helmholtz problem is satisfied
\begin{equation}
u_{n}(r\in\Omega\cup\Gamma)=(\Delta-\lambda_{n}I)^{-1}f_{n}(r\in\Gamma).\label{eq:helmholtz}
\end{equation}
This can be seen by rewriting Eq. \ref{eq:helmholtz} 
\[
\Delta u_{n}-f_{n}=\lambda_{n}u_{n}
\]
and letting the function $f_{n}$ equal

\[
f_{n}=(\frac{\partial}{\partial\hat{n}}+a)\mid_{\Gamma}u_{n}.
\]
This reviel a source term $f_{n}$ that can be viewed as inducing
the boundary conditions in equation \ref{eq:eigenequation} and is
a well explored concept in potential theory\cite{Burton1971}. For
Neumann conditions $f_{n}$ consist of dipoles and for Dirichlet conditions
it consist of monopoles. By using the identity 
\[
(Q+P)^{-1}=Q^{-1}-Q^{-1}P(Q+P)^{-1}
\]
 together with Eq. \ref{eq:helmholtz} the following integral equation
is found
\begin{equation}
u_{n}=\Delta^{-1}f_{n}+\lambda_{n}\Delta^{-1}(\Delta-\lambda_{n}I)^{-1}f_{n}.\label{eq:integral_equation}
\end{equation}
 Let us for a moment assume that the exterior boundary ($\Omega$)
is infinitely large and investigate the second term in Fourier space.
Let us denote the transform vector by 
\[
|q\rangle=|q_{x}q_{y}q_{z}\rangle=e^{i(q_{x}\hat{x}+q_{y}\hat{y}+q_{z}\hat{z})\cdot r}.
\]
We get

\begin{multline}
\lambda_{n}\intop_{-\infty}^{\infty}dq|q\rangle\langle q|\Delta^{-1}(\Delta-\lambda_{n}I)^{-1}|f_{n}\rangle=\\
=\lambda_{n}\intop_{-\infty}^{\infty}dq\frac{|q\rangle\langle q|(\Delta-\lambda_{n}I)^{-1}|f_{n}\rangle}{||q||^{2}}.\label{eq:second_term}
\end{multline}
The inverse to the Helmholtz operator in Eq. \ref{eq:second_term}
can be directly evaluated in Fourier space. A fundamental solution
to the Helmholtz operator $G(r,r_{0},\kappa)$ satisfy

\begin{equation}
(\Delta-\kappa^{2}I)G(r,r_{0},\kappa)=\delta(r-r_{0})\label{eq:fundamental_solution}
\end{equation}
in any dimension. Fourier transforming equation \ref{eq:fundamental_solution}
yield
\begin{equation}
\tilde{G}(q,r_{0},\kappa)=-\frac{e^{-ir_{0}\cdot q}}{(\kappa^{2}-||q||^{2})}\label{eq:fundamental_in_fourier}
\end{equation}
where the singularity can be avoided by a small displacement $\kappa\rightarrow(\kappa+i\epsilon)$.
It is noted that the fundamental solution is local in Fourier space
in the sence that it quickly decays when $||q||>>|\kappa|$. Our goal
is to show that this hold for a general surface distribution $|f\rangle$.
We begin by investigating two charges
\[
|f\rangle=\delta(r)+\delta(r-r_{0})
\]
and apply the steps in equation \ref{eq:fundamental_solution}-\ref{eq:fundamental_in_fourier}
leading to 
\begin{equation}
\tilde{G}(q,r_{0},\kappa)=\frac{1-e^{-ir_{0}\cdot q}}{(\kappa^{2}-||q||^{2})}\propto||q||^{-2}\mbox{ for }||q||>>|\kappa|.\label{eq:result_helmholtz_operator}
\end{equation}
Hence the result of introducing two (separated) charges merely introduces
a modulation of the solution in Fourier space, with a frequency associated
to the separation of the charges. It is thus concluded that the Fourier
expansion in equation \ref{eq:second_term} is local around $\kappa^{2}$
and that this holds regardless of the number of sources and positions
of the sources. In other words \emph{regardless of the shape of the
boundary $\Gamma$. }This also hold in the case where exterior boundary
conditions are imposed (by analysis of Fourier series). The second
term in Eq. \ref{eq:second_term} can thus be truncated to the following
expression
\begin{multline}
...\approx\lambda_{n}\intop_{-(|\lambda_{n}|+C)}^{|\lambda_{n}|+C}dq\frac{|q\rangle\langle q|(\Delta-\lambda_{n}I)^{-1}|f_{n}\rangle}{||q||^{2}}+\lambda_{n}O(\lambda_{n}^{-2}|\lambda_{n}+C|^{-2}|)\label{eq:trunk}
\end{multline}
for some scalar $C$. Eq. \ref{eq:second_term}-\ref{eq:result_helmholtz_operator}
involve the free-space Helmholtz operator. The original eigenvalueequation
is bounded in a domain, typically Neumann, Dirichlet or a periodic.
The inverse to the Helmholtz operator in such domains will also be
local in Fourier space and have the same asymptotic behavior (Neumann,
Dirichlet or periodic conditions give a subset of the free space operator
in Fourier space). The equivalent of Eq. \ref{eq:trunk} for such
domains is a truncated Fourier series expansion
\begin{equation}
\tilde{u}_{n}\approx\sum_{q=1}^{\infty}|q\rangle\langle q|\Delta^{-1}|f_{n}\rangle+\sum_{q=1}^{N}\frac{|q\rangle\langle q|\langle q|(\Delta-\lambda_{n}I)^{-1}|f_{n}\rangle}{|\lambda_{q}|}+O(q^{-2}|\lambda_{N}|^{-2})\label{eq:fourier_expanded_eigenfunction}
\end{equation}
for some $N$. On such a domain (Dirichlet) one has
\[
\delta(r)=\frac{1}{2\pi}\sum_{-\infty}^{\infty}e^{iqr}
\]
 Let us now assume that the correct surface distribution $|f_{n}\rangle$
satisfying Eq. \ref{eq:integral_equation} is unknown. This can be
formulated in the following way:

Given a boundary $\Gamma$ we seek a correct surface distribution
$|f_{n}(r\in\Gamma)\rangle$ such that when plugged in equation \ref{eq:fourier_expanded_eigenfunction}
yield the correct sought eigenfunction \uline{$u_{n}$}. This can
be solved by using a (any complete) series expansion over the boundary
$\Gamma$. A harmonic expansion is suitable~\cite{Nordin2011} (but
probably not optimal) 
\begin{equation}
|f_{n}\rangle=\sum_{\sigma=0}^{\infty}|\sigma_{n}(r\in\Gamma)\rangle\langle\sigma_{n}|f_{n}\rangle=\sum_{\sigma=0}^{\infty}\beta_{\sigma n}|\sigma_{n}\rangle.\label{eq:surface_expansion}
\end{equation}
Combining equation \ref{eq:surface_expansion} with equation \ref{eq:fourier_expanded_eigenfunction}
we get

\[
\tilde{u}_{n}\approx\sum_{\sigma=1}^{\infty}\sum_{q=1}^{\infty}\beta_{\sigma n}|q\rangle\langle q|\Delta^{-1}|\sigma\rangle+\sum_{\sigma=1}^{\infty}\sum_{q=1}^{N}\frac{\beta_{\sigma n}|q\rangle\langle q|\langle q|(\Delta-\lambda_{n}I)^{-1}|\sigma_{n}\rangle}{|\lambda_{q}|}+O(q^{-2}|\lambda_{N}|^{-2})
\]
The first term has poor convergence in Fourier space. An efficient
solution to this problem is to construct an orthogonal complement
to the set $\{|q\rangle\}_{q=1}^{N}$ using for Dirichlet conditions
on $\Gamma$ the monopole kernel
\[
|s\rangle=\intop_{\Omega}\frac{\sigma_{s}(r')\hat{n}(r')}{||r-r'||}dr'
\]
and in the case of Neumann conditions the dipole kernel 
\[
|s\rangle=\intop_{\Omega}\frac{\sigma_{s}(r')\hat{n}(r')}{||r-r'||^{2}}dr'.
\]
Then an orthogonal series can be constructed using the expansion of
the surface expansion $\{|\sigma\rangle\}_{\sigma=1}^{\infty}$ in
the following way
\begin{align*}
|s_{n1}\rangle= & \Delta^{-1}|\sigma=1_{n}\rangle.\\
|s_{nk}\rangle= & \Delta^{-1}|\sigma=k_{n}\rangle-\sum_{j=1}^{k-1}|s_{nj}\rangle\langle s_{nj}|\Delta^{-1}|\sigma=k_{n}\rangle-\sum_{j=1}^{N}|q_{j}\rangle\langle q_{j}|\Delta^{-1}|\sigma=k_{n}\rangle\\
\vdots\\
\mbox{for }k\rightarrow\infty & .
\end{align*}
By this, a mixed basis is obtained 
\begin{equation}
\{|q=1\rangle,|q=2\rangle,...,|q=N\rangle,|s=1\rangle,|s=2\rangle,...,|s\rightarrow\infty\rangle\}.\label{eq:mixed_basis}
\end{equation}
It is a straight forward excercise to show that the constructed infinite
series $\{|s_{n1}\rangle,|s_{n2}\rangle,...\}$ span a strict subspace
of the function space in the joint domain $\Omega\cup\Gamma$. In
fact $\{q\}_{q=1}^{N}\cup\{|s\rangle\}_{s=1}^{\infty}$ span approximately
the first $N$ sought eigenfunctions $u_{n}$ (with an error $O(q^{-2}|\lambda_{N}|^{-2})$).
Furthermore if one lets $N$ grow to infinity, the orthogonal set
$\{|s\rangle\}_{s=1}^{\infty}$ is forced to the null-space by its
construction i.e. the mixed basis is not over-determined (it is however
not complete, by the truncation). Therefore one may choose a truncation
$N$ and effectively capture the low-frequent behaviour of the Laplace
operator in a bounded domain.

Returning to the original problem Eq. \ref{eq:eigenequation} where
the eigenvalues $\lambda_{n}$ and eigenfunctions $u_{n}$ are unknown,
a perturbation matrix $A$ using the mixed basis in Eq. \ref{eq:mixed_basis}
can be formed which captures the relevant low-frequence information
of the eigenproblem stated in Eq. \ref{eq:eigenequation} in the following
way
\begin{equation}
A_{nm}=\begin{cases}
\langle q_{n}|\Delta|q_{m}\rangle_{\Omega}-\langle q_{n}|(\frac{\partial}{\partial\hat{n}}+a)|q_{m}\rangle_{\Gamma} & \mbox{if }n,m\leq N\\
\langle q_{n}|\Delta|s_{m}\rangle_{\Omega}-\langle q_{n}|(\frac{\partial}{\partial\hat{n}}+a)|s_{m}\rangle_{\Gamma} & \mbox{if }n\leq N<m\\
\langle s_{n}|\Delta|q_{m}\rangle_{\Omega}-\langle s_{n}|(\frac{\partial}{\partial\hat{n}}+a)|q_{m}\rangle_{\Gamma} & \mbox{if }m\leq N<n\\
\langle s_{n}|\Delta|s_{m}\rangle_{\Omega}-\langle s_{n}|(\frac{\partial}{\partial\hat{n}}+a)|s_{m}\rangle_{\Gamma} & \mbox{if }N<n,m
\end{cases}\label{eq:perturbation_matrix}
\end{equation}
where the subscript $\Gamma$ is a reminder of the fact that the inner
products are calculated on the boundaries only using Eq. \ref{eq:inner_products_boundaries}.
Let us take a moment and look at the integrals involving the boundary
conditions. The first term involving $q_{n},q_{m}$ is easily evaluated
using Eq. \ref{eq:inner_products_boundaries}. The integrals involving
the surface functions are however a bit more subtle, due to the infinities
of the potentials at the origin of the charges. We get using (Eq.
\ref{eq:inner_products_boundaries}) 
\[
\langle q_{n}|(\frac{\partial}{\partial\hat{n}}+a)|s_{m}\rangle_{\Gamma}=\intop_{\Gamma}q_{n}(r)(\hat{n}(r)\cdot\nabla+a)s_{m}(r)dr=\intop_{\Gamma}q_{n}(r)(\hat{n}(r)\cdot\nabla+a)\intop_{\Gamma}(\frac{\partial}{\partial\hat{n}}+a)\frac{\sigma_{m}(r')}{||r-r'||}dr'dr=
\]
\[
=\intop_{\Gamma}q_{n}(r)(\frac{\partial}{\partial\hat{n}}+a)\intop_{\Gamma}(\frac{\partial}{\partial\hat{n}}+a)\frac{\sigma_{m}(r')}{||r-r'||}dr'dr.
\]
Now we split the rightmost integral in two parts 
\[
\intop_{\Gamma}(\frac{\partial}{\partial\hat{n}}+a)\frac{\sigma_{m}(r')}{||r-r'||}dr'=\Phi_{0}(r)+\Phi_{1}(r)
\]
where $\Phi_{0}$ is the potential from the charges located at the
position of $r$ (self-interaction) and $\Phi_{1}$ is the potential
arising from surrounding charges. Evidently $\Phi_{1}$ is finite.
For $\Phi_{0}$ we get (in the case of Neumann) 
\[
\hat{n}\cdot\nabla\Phi_{0}\rightarrow\infty
\]
but importantly this only evaluated at $\Gamma$ (and is symmetric
across the boundary) and therefore efficiently acts as a $\delta'$-function.
Therefore 
\[
\intop_{\Gamma}q_{n}(r)\hat{n}\cdot\nabla\Phi_{0}(r)dr=C\intop_{\Gamma}\sigma_{m}(r)\hat{n}\cdot\nabla q_{n}(r)dr
\]
for some constant $C<\infty$. The $\Phi_{1}(r)$ contribution is
evaluated as 
\[
\intop_{\Gamma}q_{n}(r)\hat{n}\cdot\nabla\Phi_{1}(r)dr=\intop_{\Gamma}q_{n}(r)\hat{n}(r)\cdot\nabla\intop_{\Gamma}\hat{n}(r')\cdot\nabla\frac{\sigma_{m}(r')}{||r-r'||}dr'dr\,\mbox{ (}r\neq r')
\]
\[
=\intop_{\Gamma}q_{n}(r)\hat{n}(r)\cdot\nabla\Phi_{1}(r)=-\intop_{\Gamma}q_{n}(r)\hat{n}(r)\cdot F(r)dr
\]
\[
=-\intop_{\Gamma}drq_{n}(r)\hat{n}(r)\cdot\intop_{\Gamma}dr'\sigma_{m}(r')\left(\frac{3\hat{n}(r')\cdot\hat{a}}{||r-r'||^{3}}\hat{a}-\frac{\hat{n}(r')}{||r-r'||^{3}}\right).
\]
where $\hat{a}=(r-r')/||r-r'||$ is the unit vector pointing from
$r$ towards $r'$. which effectively gives the force due to surrounding
dipoles along the boundary $\Gamma$. This can be interpreted as calculating
the work needed to move a charged particle along $\Gamma$ subject
to the field emerging from a dipole distribution. This can be calculated
for each individual dipole and the total contribution can be found
by integration.  Note that if the surface is smooth, the function
behaves (relatively) nice, as the scalar product disappears in case
of a flat surface. The products
\[
\langle s_{n}|(\frac{\partial}{\partial\hat{n}}+a)|s_{m}\rangle_{\Gamma}
\]
are evaluated in a similar way. Together with a orthogonalization
matrix
\begin{equation}
B_{nm}=\begin{cases}
\delta_{nm} & \mbox{if }n,m\leq N\\
\langle q_{n}|s_{m}\rangle_{\Omega} & \mbox{if }n\leq N<m\\
\langle s_{n}|q_{m}\rangle_{\Omega} & \mbox{if }m\leq N<n\\
\langle s_{n}|s_{m}\rangle_{\Omega} & \mbox{if }N<n,m
\end{cases}\label{eq:orthogonalization_matrix}
\end{equation}
an approximation to the $N$ first eigenvalues and eigenfunctions
to the eigenproblem is found by diagonalizing the orthogonalization
matrix $B$
\[
VDV=B
\]
and thus forming a basis transformation $W$
\[
W=V\sqrt{D^{-1}}
\]
which can be used to diagonalize 
\[
V_{A}D_{A}V_{A}=W^{T}AW.
\]
$D_{A}$ then contain an approximation to the $N$ first eigenvales
of the eigenproblem, and an approximation to the corresponding $N$
first eigenfunctions (in the mixed basis) can be read out by the columns
of $V_{A}$. Furthermore, in the low-frequent domain the eigenfunctions
are expected to vary slowly also locally to the boundaries. Therefore
in practice, the surface expansion $\{|\sigma\rangle\}_{\sigma=1}^{\infty}$
can be truncated to a finite value $\{|\sigma\rangle\}_{\sigma=1}^{M}$.
In fact, this value is expected to be quite low since the variation
of all possible surface modes $|\sigma\rangle$ on $\Gamma$ is smeared
out to small variations in the volume $\Omega$ by the integral operator
$\Delta^{-1}$. The analysis of the trunctation of the surface expansion
is left out in this study with the comment that previous numerical
studies show that in practice a low number $M$ yield good results~(see
e.g. \cite{Nordin2011}). Previously~\cite{Nordin2011} it has also
been shown that all inner producs in \ref{eq:perturbation_matrix}-\ref{eq:orthogonalization_matrix}
can be transformed to surface integrals using the self-adjointness
of the Laplace operator and the following two relations

\begin{align*}
\Delta|q\rangle= & \lambda_{q}|q\rangle\\
\Delta|s\rangle= & |\sigma_{s}\rangle.
\end{align*}
A few more useful insights are reported.For Neumann conditions the
charge distributions $|\sigma\rangle$ consist of dipole distributions.
In this case the inner product of the resulting potentials appearing
in Eq. \ref{eq:orthogonalization_matrix} can be evaluated on the
surface in the following way 
\begin{equation}
\langle s_{n}|s_{m}\rangle=\begin{cases}
\intop_{\Gamma}\intop_{\Gamma}drdr'\sigma_{n}(r)\sigma_{m}(r')\frac{\hat{n}_{\perp}(r)\cdot\hat{n}_{\perp}(r')}{||r-r'||} & \mbox{In the case of Dirichlet conditions on }\mbox{\ensuremath{\Gamma}}\\
\intop_{\Gamma}\intop_{\Gamma}drdr'\sigma_{n}(r)\sigma_{m}(r')\frac{\hat{n}(r)\cdot\hat{n}(r')}{||r-r'||} & \mbox{In the case of Neumann conditions on }\Gamma.
\end{cases}\label{eq:surface_kernel}
\end{equation}
A derivation of this result is attached as an appendix of this paper.
\begin{multline}
\mid s\rangle=\intop_{\Omega}\frac{\sigma_{s}(r')\hat{n}(r')}{||r-r'||^{2}}dr'\Rightarrow\\
\langle s\mid x\mid s'\rangle=\int xdr\intop_{\Omega}\frac{\sigma_{s}(r')\hat{n}(r')}{||r-r'||^{2}}dr'\intop_{\Omega}\frac{\sigma_{s'}(r'')\hat{n}(r'')}{||r-r''||^{2}}dr''=\\
\intop_{\Omega}\sigma_{s}(r')\Xi_{g}(r',r'')\sigma_{s}(r')dr'dr''.\label{eq:inner_product_xi_kernel}
\end{multline}

\begin{multline}
\Xi=\int dr''(g\cdot r'')\frac{\hat{n}(r)}{||r''-r||^{2}}\frac{\hat{n}(r')}{||r''-r'||^{2}}=\\
\pi[g\times\hat{a}(r,r')]\cdot[\hat{n}(r)\times\hat{n}(r')]\\
-\pi g\cdot\hat{a}(r,r')\left([\hat{n}(r)\cdot\hat{n}(r')]\right.\\
\left.-[\hat{n}(r)\cdot\hat{a}(r,r')][\hat{n}(r')\cdot\hat{a}(r,r')]\right)\label{eq:xi_kernel}
\end{multline}
where $\hat{a}=\frac{r-r'}{||r-r'||}$ is the directional (unit) vector
between $r$ and $r'$ and $\hat{n}$ is the (outward) pointing normal
at the boundary $\Omega$. A few comments on the result are appropriate.
If the eigenbasis of the kernel appearing in Eq. \ref{eq:surface_kernel}
is found, orthogonal surface functions can be constructed directly.
This is realized by letting the set $\{\sigma_{n}\}_{n=1}^{\infty}$
equal the eigenbasis of $\Theta$ i.e. $\langle\sigma_{s}|\Theta|\sigma_{s'}\rangle=\delta_{ss'}\lambda_{s'}$
where $\lambda_{s'}$ equal the eigenvalue of $\Theta$ corresponding
to the eigenfunction $|\sigma_{s}\rangle$. In example: A special
case is found when the surface is flat (independent of the scalar
product between the surface normals). The kernel $\Theta$ then reduces
to the Poisson kernel and the eigenfunctions are analytically known.
Similar simple expressions are expected from simple domains such as
spheres, cylinders et cetera. Importantly this introduces the possibility
of filling a space with such bodies and solve approximately the eigenvalueproblem
as well as the Bloch-Torrey equation in the void space between the
bodies in a mesh-free way with the above approach. The demanding computational
step is then to find the electrostatic potentials between such bodies.
This is a standard problem and solutions using fast multipole methods
have been proposed~\cite{Darve2000,Greengard1998,jiang2005}. Furthermore
non-trivial bodies could be approximated by discrete grids. This would
yield a finite set of surface functions. A harmonic expansion on such
surfaces could easily be constructed by mapping the spherical harmonics
to such bodies and it is expected that few such functions are needed
for good results. It has been shown that such calculations can be
performed in optimal time with respect to the number of discretation
points on the surfaces. On a computational note, the kernel in Eq.
\ref{eq:surface_kernel} is symmetric and problem independent. It
can thus be calculated off-line and in an implementation interpolation
can be made using the distance between the surface elements and the
scalar product between the surface normals.

\section*{Appendix: Derivation of the dipole kernel}

For the mixed basis to make sense, it must be orthogonalized before
the perturbation matrix is formed (for details, see~\cite{Nordin2009,Nordin2011})
. In particular this require the inner product between the surface
functions $\langle s|s'\rangle$ which are defined throughout the
whole volume. These inner products can be transformed to surface integrals
and here follows a derivation of this result. The surface functions
are defined as solutions to the inhomogeneous Poisson's equation
\[
|s\rangle=\intop_{\Omega}\frac{\sigma_{s}(r')\hat{n}(r')}{||r-r'||^{2}}dr'
\]
where $\sigma_{s}$ denote a dipole distribution at the surface and
$\hat{n}$ the (outward) pointing surface normal and the kernel is
the fundamental solution for a dipole potential~ \cite{Jackson1998}.
Typically we want to express the surface contributions of $S$ by
a function expansion over the surface. Since the $S$-operator is
located to the boundaries and thus has a huge null-space more or less
any truncated function expansion on the surface will capture the low
frequency part of the (volume) contribution. The reason for this is
that the null-space is known trivially. A Fourier expansion on the
surface is suggested as the low frequency part is well captured by
the first $M$ Fourier functions on the surface. The (volume) inner
products are formally written as
\begin{equation}
\langle s_{n}(r)|s_{m}(r)\rangle=\int\intop_{\Omega}\frac{\hat{n}(r')\cdot r\mathbf{\mathbf{\sigma_{n}}}(r')}{||r-r'||^{2}}dr'\intop_{\Omega}\frac{\hat{n}(r'')\cdot r\mathbf{\mathbf{\sigma_{m}}}(r'')}{||r-r''||^{2}}dr''dr.\label{eq:inner_product_potentials}
\end{equation}
Where the outer integral is the volume integral. The convergence of
the Poisson dipole kernel ensure us that we can interchange the order
of integration, this results in

\begin{multline*}
=\iintop_{\Omega}\intop_{V}\frac{\hat{n}(r')\cdot r\mathbf{\mathbf{\sigma}}(r')}{||r-r'||^{2}}\frac{\hat{n}(r'')\cdot r\mathbf{\mathbf{\sigma}}(r'')}{||r-r''||^{2}}dVdr'dr''=\\
\iintop_{\Omega}\mathbf{\mathbf{\sigma}}(r')\mathbf{\mathbf{\sigma}}(r'')\Theta(r',r'')\hat{n}(r')\cdot\hat{n}(r'')dr'dr''
\end{multline*}
where the kernel $\Theta$ is defined as
\[
\Theta(r',r'')=\intop_{V}\frac{\hat{n}(r')\cdot r}{\text{||}r-r'||^{2}}\frac{\hat{n}(r'')\cdot r}{||r-r''||^{2}}dV.
\]

A multipole expansion of the potential from a dipole located in origo
can be written as~\cite{Jackson1998} 
\begin{equation}
\Phi_{0}(x)=\frac{1}{4\pi\epsilon_{0}}\sum_{l=0}^{\infty}\sum_{m=-l}^{l}\frac{4\pi}{2l+1}q_{lm}^{0}\frac{Y_{lm}(\theta,\phi)}{r^{l+1}}.\label{eq:multipole}
\end{equation}
Let us denote the dipole by $p_{0}=\hat{n}(0)d$ (and ensure that
it is charge neutral). The scalar factors in equation \ref{eq:multipole}
are found by 
\begin{equation}
q_{lm}^{0}=\int Y_{lm}^{*}(\theta,\phi)r^{l}\rho_{0}(\mathbf{x})d^{3}x\label{eq:dipole_moments}
\end{equation}
where $\rho_{0}$ denotes the charge distribution. The only surviving
terms in equation \ref{eq:dipole_moments} are the dipole moments

\[
\begin{aligned}q_{11}^{0}= & -\sqrt{\frac{3}{8\pi}}(p_{0x}-ip_{0y})\\
q_{10}^{0}= & \sqrt{\frac{3}{4\pi}}p_{0z}\\
q_{1-1}^{0}= & -q_{11}^{*}
\end{aligned}
\]
where $i$ denote the imaginary unit. For the second dipole located
in the $z$-axis $p_{1}=$$\hat{n}(a\hat{z})d$ not only the dipole
moment survives but also higher modes. The potential from this dipole
$\Phi_{1}(x)$ can still be expanded around origo as 

\[
\begin{aligned}q_{00}^{1}= & 0\\
q_{11}^{1}= & -\sqrt{\frac{3}{8\pi}}2d\sin\theta_{1}[\cos\phi_{1}-i\sin\phi_{1}]\\
q_{10}^{1}= & \sqrt{\frac{3}{4\pi}}2d\cos\theta_{1}\\
q_{22}^{1}= & 0\\
q_{21}^{1}= & -\sqrt{\frac{15}{8\pi}}2ad\sin\theta_{1}[\cos\phi_{1}-i\sin\phi_{1}]\\
q_{20}^{1}= & \frac{1}{2}\sqrt{\frac{5}{4\pi}}8ad\cos\theta_{1}\\
\vdots
\end{aligned}
.
\]

\[
\]
We are interested in calculating 
\[
\langle\Phi_{0},\Phi_{1}\rangle=\intop_{V}\Phi_{0}(r)\Phi_{1}(r)dr=...
\]
and note that by the orthogonality of the spherical harmonics that
the only surviving modes are the dipole modes 
\[
...=(\frac{1}{3\epsilon_{0}})^{2}[-q_{11}^{0}q_{1-1}^{1}+q_{10}^{0}q_{10}^{1}-q_{1-1}^{0}q_{11}^{1}]\intop_{a}^{\infty}\frac{1}{r^{4}}r^{2}dr=
\]
 by evaluating the radial integral we get 
\[
=p_{0}\cdot p_{1}\frac{1}{a}.
\]
Since $a$ is the distance between the two dipoles (and hence positive)
we conclude

\[
\langle\Phi_{0},\Phi_{1}\rangle=\frac{\mathbf{p_{0}}\cdot\mathbf{p_{1}}}{||\mathbf{r_{0}}-\mathbf{r_{1}}||}=\frac{\hat{n}(r')\cdot\hat{n}(r'')}{||r'-r''||}=\Theta(r',r'').
\]
Therefore, the (volume) inner product of two potentials $\langle s|s'\rangle$
in equation \ref{eq:inner_product_potentials} can be reduced to a
(double) surface integral 
\[
\langle s|s'\rangle=\intop_{\Omega}\sigma_{s}(r')\sigma(r'')\Theta(r',r'')dr'dr''.
\]
A few comments on the result are appropriate. First, the kernel $\Theta$
is symmetric and problem independent, it can thus be calculated off-line
and in an implementation interpolation can be made using the distance
between the surface elements and the scalar product between the surface
normals. Secondly, this type of kernels can be approximated by single
integrals using multipole methods. This has not yet been tested. Furthermore,
if the eigenbasis of the kernel $\Theta$ is found, orthogonal bases
can be constructed directly. This is realized by letting the set $\{\sigma_{s}\}_{s=1}^{M}$
equal the eigenbasis of $\Theta$ i.e. $\langle\sigma_{s}|\Theta|\sigma_{s'}\rangle=\delta_{ss'}\lambda_{s'}$
where $\lambda_{s'}$ equal the eigenvalue of $\Theta$ corresponding
to the eigenfunction $|\sigma_{s}\rangle$. In example: A special
case is found when the surface is flat (independent of the scalar
product between the surface normals). The kernel $\Theta$ then reduces
to the Poisson kernel and the eigenfunctions are analytically known.

\bibliographystyle{unsrt}

\begin{thebibliography}{10}

\bibitem{Stejskal1965a}
E.~O. Stejskal and J.~E. Tanner.
\newblock {Spin Diffusion Measurements: Spin Echoes in the Presence of a
  Time-Dependent Field Gradient}.
\newblock {\em The Journal of Chemical Physics}, 42(1):288, 1965.

\bibitem{callaghan1991book}
P.~T. Callaghan.
\newblock {\em Principles of Nuclear Magnetic Resonance Microscopy}.
\newblock Oxford University Press, 1991.

\bibitem{price2009}
William~S Price.
\newblock {\em {NMR Studies of Translational Motion}}.
\newblock Cambridge University Press, 2009.

\bibitem{Mitra1992}
Partha Mitra, Pabitra Sen, Lawrence Schwartz, and Pierre {Le Doussal}.
\newblock Diffusion propagator as a probe of the structure of porous media.
\newblock {\em Physical Review Letters}, 68(24):3555--3558, June 1992.

\bibitem{Mitra1993}
Partha~P. Mitra.
\newblock Short-time behavior of the diffusion coefficient as a geometrical
  probe of porous media.
\newblock {\em Physical Review B}, 47(14):8565--8574, April 1993.

\bibitem{Hurlimann1994}
M.~H\"{u}rlimann.
\newblock Restricted diffusion in sedimentary rocks. determination of
  surface-area-to-volume ratio and surface relaxivity.
\newblock {\em Journal of Magnetic Resonance, Series A}, 111(2):169--178, 1994.

\bibitem{callaghan1991}
Paul~T Callaghan.
\newblock {\em {Principles of Nuclear Magnetic Resonance Microscopy}}.
\newblock Oxford University Press, 1991.

\bibitem{topgaard2003}
D.~Topgaard and O.~S\"{o}derman.
\newblock {Experimental determination of pore shape and size using q-space NMR
  microscopy in the long diffusion-time limit.}
\newblock {\em Magnetic resonance imaging}, 21:69--76, 2003.

\bibitem{Malmborg2006}
C.~Malmborg, M.~Sj\"{o}beck, S.~Brockstedt, E.~Englund, O.~S\"{o}derman, and
  D.~Topgaard.
\newblock {Mapping the intracellular fraction of water by varying the gradient
  pulse length in q-space diffusion MRI.}
\newblock {\em Journal of Magnetic Resonance}, 180(2):280--5, 2006.

\bibitem{Price2003}
William~S Price, Peter Stilbs, and Olle S\"{o}derman.
\newblock Determination of pore space shape and size in porous systems using
  nmr diffusometry. beyond the short gradient pulse approximation.
\newblock {\em Journal of Magnetic Resonance}, 160(2):139--143, 2003.

\bibitem{Latour1995}
L.~L. Latour.
\newblock {Pore-Size Distributions and Tortuosity in Heterogeneous Porous
  Media}.
\newblock {\em Journal of Magnetic Resonance, Series A}, 112(1):83--91, 1995.

\bibitem{Mair2000}
R.~W. Mair, M.~D. H\"{u}rlimann, P.~N. Sen, L.~M. Schwartz, S.~Patz, and R.~L.
  Walsworth.
\newblock {Tortuosity measurement and the effects of finite pulse widths on
  xenon gas diffusion NMR studies of porous media.}
\newblock {\em Magnetic resonance imaging}, 19(3-4):345--51, 2000.

\bibitem{Mair2002}
R.~W. Mair, P.~N. Sen, M.~D. H\"{u}rlimann, S.~Patz, D.~G. Cory, and R.~L.
  Walsworth.
\newblock {The narrow pulse approximation and long length scale determination
  in xenon gas diffusion {NMR} studies of model porous media.}
\newblock {\em Journal of Magnetic Resonance}, 156(2):202--12, 2002.

\bibitem{Grebenkov}
D.~Grebenkov.
\newblock Use , misuse , and abuse of apparent diffusion coefficients.
\newblock {\em Concepts in Magnetic Resonance}, pages 24--35, 2010.

\bibitem{Grebenkov2007}
Denis Grebenkov.
\newblock Nmr survey of reflected brownian motion.
\newblock {\em Reviews of Modern Physics}, 79(3):1077--1137, August 2007.

\bibitem{torrey1956}
H~C Torrey.
\newblock {Bloch Equations with Diffusion Terms}.
\newblock {\em Phys. Rev.}, 104(3):563--565, November 1956.

\bibitem{Nordin2011}
Matias Nordin, Martin Nilsson-Jacobi, and Magnus Nyd\'{e}n.
\newblock A mixed basis approach in the {SGP}-limit.
\newblock {\em Journal of magnetic resonance (San Diego, Calif. : 1997)},
  212(2):274--9, October 2011.

\bibitem{linse1995}
P.~Linse and O.~S\"{o}derman.
\newblock The validity of the short-gradient-pulse approximation in {NMR}
  studies of restricted diffusion. simulations of molecules diffusing between
  planes, in cylinders and spheres.
\newblock {\em Journal of Magnetic Resonance, Series A}, 116(1):77--86, 1995.

\bibitem{Barzykin1999}
A~V Barzykin.
\newblock Theory of spin echo in restricted geometries under a step-wise
  gradient pulse sequence.
\newblock {\em Journal of Magnetic Resonance}, 139(2):342--353, 1999.

\bibitem{Kenkre1997}
V~M Kenkre, Eiichi Fukushima, and D~Sheltraw.
\newblock Simple solutions of the {T}orrey-{B}loch equations in the {NMR} study
  of molecular diffusion.
\newblock {\em Journal of Magnetic Resonance}, 128(1):62--69, 1997.

\bibitem{Amini1999}
S.~Amini.
\newblock {On boundary integral operators for the Laplace and the Helmholtz
  equations and their discretisations}.
\newblock {\em Engineering Analysis with Boundary Elements}, 23(4):327--337,
  April 1999.

\bibitem{Bonnet1998}
Marc Bonnet, Cnrs Ura, Ecole Polytechnique, Palaiseau Cedex, and Giulio Maier.
\newblock {Symmetric Galerkin boundary element method}.
\newblock {\em Appl. Mech. Rev.}, 51:669--704, 1998.

\bibitem{Chen2007}
J.T. Chen, C.S. Wu, Y.T. Lee, and K.H. Chen.
\newblock {On the equivalence of the Trefftz method and method of fundamental
  solutions for Laplace and biharmonic equations}.
\newblock {\em Computers \& Mathematics with Applications}, 53(6):851--879,
  March 2007.

\bibitem{Burton1971}
A.~J. Burton and G.~F. Miller.
\newblock {The Application of Integral Equation Methods to the Numerical
  Solution of Some Exterior Boundary-Value Problems}.
\newblock {\em Proceedings of the Royal Society A: Mathematical, Physical and
  Engineering Sciences}, 323(1553):201--210, June 1971.

\bibitem{Darve2000}
Eric Darve.
\newblock The fast multipole method: Numerical implementation.
\newblock {\em Journal of Computational Physics}, 160(1):195 -- 240, 2000.

\bibitem{Greengard1998}
Leslie Greengard, Jingfang Huang, Vladimir Rokhlin, and Stephen Wandzura.
\newblock Accelerating fast multipole methods for the helmholtz equation at low
  frequencies.
\newblock {\em IEEE Comput. Sci. Eng.}, 5(3):32--38, July 1998.

\bibitem{jiang2005}
Li~Jun Jiang and Weng~Cho Chew.
\newblock A mixed-form fast multipole algorithm.
\newblock {\em Antennas and Propagation, IEEE Transactions on}, 53(12):4145 --
  4156, dec. 2005.

\bibitem{Nordin2009}
Matias Nordin, Martin Nilsson-Jacobi, and Magnus Nyd\'{e}n.
\newblock {A mixed basis approach to approximate the spectrum of Laplace
  operator, Paper Presented at the VI's Proceedings of Interdisciplinary
  Transport Phenomena, Volterra, Italy,2009. arXiv:0909.0935v1.}

\bibitem{Jackson1998}
John~D. Jackson.
\newblock {\em {Classical Electrodynamics Third Edition}}.
\newblock Wiley, third edition, August 1998.

\end{thebibliography}

\end{document}